\documentclass[sn-mathphys,Numbered]{sn-jnl}

\usepackage{graphicx}%
\usepackage{multirow}%
\usepackage{amsmath,amssymb,amsfonts}%
\usepackage{amsthm}%
\usepackage{mathrsfs}%
\usepackage[title]{appendix}%
\usepackage{xcolor}%
\usepackage{textcomp}%
\usepackage{manyfoot}%
\usepackage{booktabs}%
\usepackage{algorithm}%
\usepackage{algorithmicx}%
\usepackage{algpseudocode}%
\usepackage{listings}%
\usepackage{tabularx, multirow}
\usepackage{caption}
\usepackage{longtable}
\usepackage{geometry}

\theoremstyle{thmstyleone}%
\newtheorem{theorem}{Theorem}
\newtheorem{proposition}[theorem]{Proposition}%

\theoremstyle{thmstyletwo}%
\newtheorem{example}{Example}%
\newtheorem{remark}{Remark}%

\theoremstyle{thmstylethree}%
\newtheorem{definition}{Definition}%

\begin{document}

\title[Article title]{Towards Carbon Transparency: A High-Resolution Carbon Emissions Database for China's Listed Companies}


\author[1]{\fnm{Xinlei} \sur{Wang}}\email{xinlei.wang@sydney.edu.au}
\author*[2,3]{\fnm{Junhua} \sur{Zhao}}\email{zhaojunhua@cuhk.edu.cn }
\author[4]{\fnm{Haifeng} \sur{Wu}}\email{wuhaifeng@cuhk.edu.cn}
\author[2,3]{\fnm{Zhengwen} \sur{Zhang}}\email{zhengwenzhang@link.cuhk.edu.cn}
\author[2,3]{\fnm{Guolong} \sur{Liu}}\email{liuguolong@cuhk.edu.cn}
\author[2,3]{\fnm{Wenxuan} \sur{Liu}}\email{liuwenxuan@cuhk.edu.cn}
\author[3]{\fnm{Yuheng} \sur{Cheng}}\email{yuhengcheng@link.cuhk.edu.cn}
\author[1]{\fnm{Jing} \sur{Qiu}}\email{jeremy.qiu@sydney.edu.au}
\author[4,5]{\fnm{Bohui} \sur{Zhang}}\email{bohuizhang@cuhk.edu.cn }
\author[2,3]{\fnm{Jianwei} \sur{Huang}}\email{jianweihuang@cuhk.edu.cn}

\affil*[1]{\orgdiv{School of Electrical and Information Engineering}, \orgname{The University of Sydney}, \orgaddress{\street{Camperdown}, \city{Sydney}, \postcode{2050}, \state{NSW}, \country{Australia}}}

\affil[2]{\orgdiv{Center for Crowd Intelligence}, \orgname{Shenzhen Institute of Artificial Intelligence and Robotics for Society (AIRS)}, \orgaddress{\street{Longgang District}, \city{Shenzhen}, \postcode{518129}, \state{Guangdong}, \country{China}}}

\affil[3]{\orgdiv{School of Science and Engineering}, \orgname{ The Chinese University of Hong Kong}, \orgaddress{\street{Longgong District}, \city{Shenzhen}, \postcode{ 518172}, \state{Guangdong}, \country{China}}}

\affil[4]{\orgdiv{Shenzhen Finance Institute}, \orgname{ The Chinese University of Hong Kong}, \orgaddress{\street{Longgong District}, \city{Shenzhen}, \postcode{ 518172}, \state{Guangdong}, \country{China}}}


\affil[5]{\orgdiv{School of Management and Economics}, \orgname{ The Chinese University of Hong Kong}, \orgaddress{\street{Longgong District}, \city{Shenzhen}, \postcode{ 518172}, \state{Guangdong}, \country{China}}}

\abstract{The dual-carbon goals of China necessitate precise accounting of company carbon emissions, vital for green development across all industries. Not only the company itself but also financial investors require accurate and comprehensive company-level emissions data for climate risk management. This paper introduces the structure and methodology of the High-resolution Database for Carbon Emissions of China-listed companies, integrating three primary data sources: self-disclosed environmental data from listed companies, long-accumulated national power emission data, and regional high-precision emission data derived from multi-source satellites. The database's innovation lies in the employment of artificial intelligence (AI) algorithms to aggregate multi-source satellite data. This approach enables the precise identification of carbon emission sources and the prediction of company-level carbon emissions. Consequently, this methodology robustly cross-validates self-reported direct emissions, enhancing the accuracy and granularity of company-level emission records. 
Central to the database's utility includes the provision of high-resolution company carbon emission data, which is not only highly accurate but also instrumental in carbon management and emission market transactions. By offering a more nuanced and verifiable picture of company emissions, the database supports China's broader efforts to meet its ambitious dual-carbon targets and transition towards a more sustainable and environmentally responsible economy. }




\maketitle

\begin{figure}[h]%
\centering
\includegraphics[width=1\textwidth]{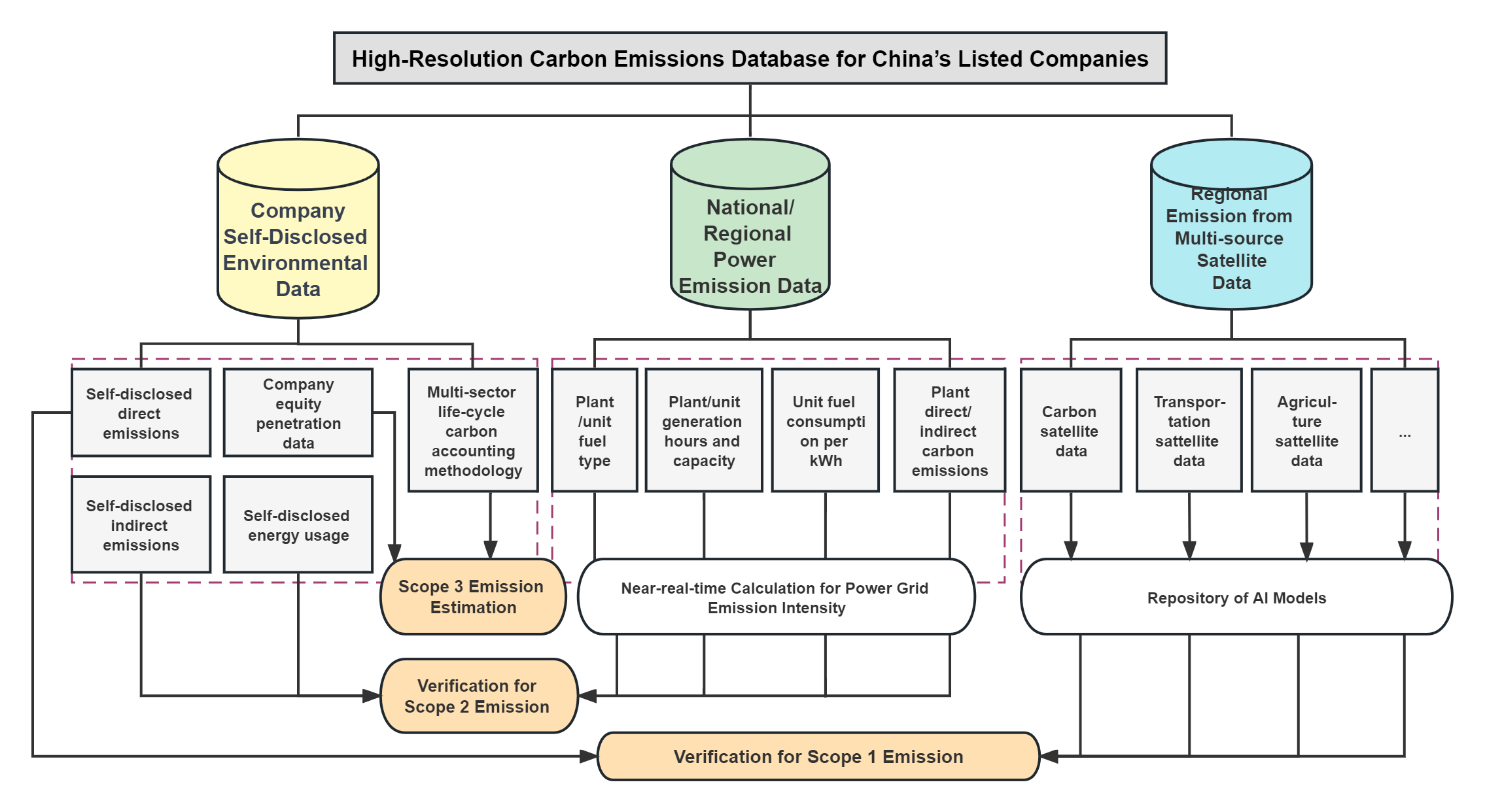}
\caption{Overview of High-Resolution Carbon Emissions Database for China's Listed Companies}\label{fig1}
\end{figure}

\section{Introduction}\label{sec1}
The adoption of China's dual carbon goal—aiming to reach peak carbon emissions by 2030 and achieve carbon neutrality by 2060—sets more stringent requirements for green development, elevating the sustainability standards across all industries within the country. Companies need precise carbon emission data accounting as an important carrier to achieve the dual carbon goal. An increasing number of investors and researchers realize that climate risk will significantly impact the financial status of listed companies. Therefore, it is strongly needed for more accurate emission information to assist investment decisions. On the one hand, real and accurate carbon emission data is the basis for the national emission trading market to carry out transactions. Thus, the foundation for company carbon accounting should be transparent, public, and complete. On the other hand, high-quality disclosure of carbon emission information can help guide capital flows, allow the public to understand the real greenhouse gas emissions of the company, and assist in achieving carbon peaking and carbon neutrality.

In the current landscape, the public disclosure of carbon emission data from companies in China's market remains inconsistent and incomplete. Particularly among A-share listed companies, the quality of the information provided varies widely.  Existing databases on company-level emissions often lack transparency in their methodologies, hindering users from understanding or replicating the data processes\cite{christensen2015incentives}. The available data is predominantly based on limited self-disclosed information, covering only a handful of companies. Concerns arise regarding the data's completeness, accuracy, and reliability\cite{kolk2008corporate}. Diverse reporting standards across companies further complicate cross-company comparisons\cite{cho2015organized}. Geographic and sectoral biases in some databases can distort broader analyses\cite{plumlee2015voluntary}. The inconsistency underscores a notable gap in the market due to a distinct lack of a comprehensive, company-level carbon emissions database that can serve as a reliable reference for stakeholders.

To address this gap, the High-Resolution Carbon Emissions Database for China’s Listed Companies is proposed in this study, incorporating multi-source satellite and power emission data into a carbon emission database. This database is mainly constructed based on three data sources:
\begin{enumerate}
    \item \textbf{Self-disclosed Environmental Data from Chinese Listed Companies}: Company's self-disclosed environmental data sourced from annual reports, Environmental, Social, and Governance (ESG) reports, social responsibility documents, official websites, and public press.
    \item \textbf{National/Regional Power Emission and Generation Data}: National/regional power emission data and power generation data released by the Chinese Government and major power companies over many years.
    \item \textbf{Regional High-precision Emission Data Combined with Multi-source Satellites}: An enhanced, spatially detailed, comprehensive, and accurate view of regional emissions from multi-source satellites.
\end{enumerate}
Satellite data provides high-resolution and accurate measurements that eliminate the biases of self-disclosed Scope 1 emissions, while power data delves into a company's Scope 2 emissions, detailing its energy consumption mix. They work together to provide a full view of a company's emissions, establishing a standard against which self-reported emission data can be validated, thereby reducing discrepancies from varying reporting standards. The global scope of satellite data ensures uniform tracking across regions and industries. Integrating the multi-source data offers a solution to traditional single-source databases' limitations by enhancing data accuracy and industry coverage range.

As Figure \ref{fig1} shows, the database integrates and cross-validates a combination of self-disclosed information and multi-source calculated data, significantly improving the accuracy and granularity of company-level emission data. Through the integration of data from multiple sources, our aim is to create a high-precision, multi-industry, and full-production-cycle company-level emissions database. The methodology for achieving this goal is outlined as follows: 

\begin{enumerate}
\item 
\textbf{Scope 1 Emissions Verification:} Utilizing the direct emissions data disclosed by the company, company equity penetration data, and the company's geographic coordinates, high-precision emissions within the company's subsidiaries and their factory area are calculated using multi-source satellite data. By comparing these two datasets, the company's Scope 1 emissions (direct emissions from owned or controlled sources) are thoroughly cross-verified; 
\item \textbf{Scope 2 Emissions Verification:} Drawing from companies' self-disclosed indirect emissions and energy consumption data, a more precise emission factor for the power grid region and indirect emissions from companies are computed based on power generation data. This subsequently facilitates the cross-validation of the company's Scope 2 emissions.; 
\item \textbf{Scope 3 Emissions Estimation:} By analyzing company equity penetration data and employing multi-industry, full-cycle carbon measurement methodologies, the Scope 3 emissions (all other indirect emissions that occur in a company's value chain) of the industry are thoughtfully estimated.
\end{enumerate}

The current public release of the High-Resolution Carbon Emissions Database for China’s Listed
Companies cover the period from January 2019 to January 2023. It includes environmental data of 5,226 A-share listed companies drawn from nearly 30,000 documents of self-disclosed reports or estimated through multi-source satellite and power generation emissions. Moreover, our research team is actively
working to extend the data for future updates.

\section{Contributions}\label{sec2}
\begin{enumerate}
\item \textbf{High-resolution Company-level Emission Estimation based on Multi-source Satellite and Power Data}: The current market is constrained by the low granularity of company-level carbon emission data, non-uniform data standards, and lagging data updates. There is a lack of internationally recognized and comprehensive company-level carbon emission databases. The carbon emissions database for listed companies in China will integrate four major data sources (self-disclosed environmental data from listed companies in China, national electricity emission data,  regional emission data from carbon satellites, and other multi-source satellites). It precisely locates the sources of company carbon emissions, monitors the carbon emissions of Chinese companies, and provides high-resolution Scope 1 and Scope 2 carbon emission data for companies. This approach not only ensures a more up-to-date snapshot but also extends coverage across a broader range of industries, thereby significantly enhancing the reliability of the data.

\item \textbf{Data Validation of Emission Estimation Results}: The company-level carbon emission database will re-verify the Scope 1 and Scope 2 emission values by integrating and cross-verifying the self-disclosed data from companies and other multi-source calculated data. Blending satellite measurements with power data analytics establishes a reference standard against which self-disclosed data can be validated. Additionally, for companies not disclosing emissions, satellite-based estimates fill the data gaps, making this approach both innovative and comprehensive. This significantly enhances the accuracy and scientific measurement of company emission disclosure data.

\item \textbf{Supporting Company-level Carbon Management and Carbon Market Trading}: By utilizing big data analysis on company electricity consumption and carbon emissions, our database provides profound insights into a company's production and operational patterns. This analysis allows us to discern the inherent characteristics of a company's power consumption, enabling real-time carbon emission accounting for companies. Through the use of open platforms and visualization tools, users can intuitively access and view company carbon emission data. This not only facilitates the analysis and comparison of carbon emissions across different companies, industries, and regions but also empowers companies in their carbon management efforts. By providing these insights, we assist companies in making informed carbon trading decisions, ultimately helping them achieve their carbon neutrality goals.
\end{enumerate}

\section{Data Collection}\label{sec0}
The data collection started from Jan 2019, mainly focusing on three main categories of data sources: self-disclosed environmental data from all China's A-share listed companies, national power emission data released by the Chinese Government and major power companies over many years, and regional high-precision emission data combined with multi-source satellites. These datasets are the fundamental input for training satellite-based Carbon Emission AI Estimation Model and validating company-level Scope 1, 2, and 3 emissions.
 
\begin{table}[!h]
\centering
\begin{tabularx}{\textwidth}{lX}
\toprule
\textbf{Disclosure Type} & \textbf{Specific Data} \\
\midrule
\multirow{5}{*}{Company Self-disclosed Emission Data} & Scope 1 and Scope 2 carbon emissions (Unit: tons) \\
 & Scope 3 emissions (Unit: tons) \\
 & Total greenhouse gas emissions of the company (Unit: tons) \\
 & Carbon emissions reporting covers the entire company or main business units \\
 & Disclosed annual emission reductions (Unit: tons) \\
\midrule
\multirow{3}{*}{Indirect Energy Usage (Scope 2 Emission Calculation)} & Company's purchased electricity (Unit: MWh) \\
 & Company's use of renewable energy power generation (Unit: MWh) \\
 & Company's purchased heat (Unit: MWh) \\
\midrule
\multirow{4}{*}{Direct Energy Usage (Scope 1 Emission Calculation)} & Company's natural gas usage (Unit: 10,000 cubic meters) \\
 & Company's coal usage (Unit: tons) \\
 & Company's gasoline usage (Unit: liters) \\
 & Company's diesel usage (Unit: liters) \\
\midrule
\multirow{2}{*}{Emissions Data for Specific Industries} & Finance: Disclosed carbon reduction loans and other green finance emission reductions (Unit: tons) \\
 & IT: Disclosed company data center emissions (Unit: tons) \\
\midrule
Equity Penetration Data & Specific equity information of the listed company's first-level subsidiaries \\
\midrule
\multirow{2}{*}{Geographic Information} & Geographic location of all subsidiaries \\
 & Geographic coordinates of all subsidiaries\\
\bottomrule
\end{tabularx}
\caption{Data Type of Company's Self-disclosed Environmental Data}\label{table1}
\end{table}

\subsection{Self-disclosed Environmental Data from China's A-share Listed Companies:}\label{subsec1}
To collect the company's self-disclosed environmental data, our primary data sources were the official websites, public documents, or press releases of all A-share listed companies and Chinese Stock Exchanges. We employed a two-pronged strategy focused on web scraping and optical character recognition (OCR) to aggregate this data from China's A-share listed corporations. We developed a comprehensive four-step process to gather self-disclosed environmental data from China's A-share listed companies, including the most recent carbon emission information.

\begin{enumerate}
\item \textbf{Data Retrieval}: From January 1, 2019, to January 1, 2023, we obtained download links for all ESG reports and annual reports—documents of A-share companies containing intricate details concerning a corporation's carbon emissions and other environmental practices. This was achieved by accessing the official websites of the Shenzhen and Shanghai Stock Exchanges and utilizing Python crawler scripts to automate the process. 
\item \textbf{Carbon Disclosure Extraction}: We then focused on extracting carbon disclosure information from these reports. This involved analyzing machine-generated PDFs that employ wireframe complete forms specifically designed for carbon information disclosure. OCR technologies are utilized to discern and extract pertinent data. We converted the graphical representations of textual data into machine-encoded text by processing these image-based files and extracted the carbon disclosure information from reports. 
\item \textbf{Algorithm Design}: To ensure accuracy in identifying and extracting the carbon disclosure information, we designed a rule-based algorithm. This algorithm takes the paths to PDF files of ESG reports and selected keywords (listed in Table \ref{table1}) as inputs. This metamorphosis facilitated a seamless extraction of valuable metrics, particularly those related to carbon emissions and other environmental indicators. 
\item \textbf{Analysis and Interpretation}: (Optional step if further analysis is conducted) We then analyzed and interpreted the extracted data to provide insights into the carbon emission trends and practices of the A-share listed companies.
\end{enumerate}


Through the above methods, we concluded and collected six categories of the company's self-disclosed environmental data. The field name, data source, and frequency of the data are as follows:
\begin{itemize}
    \item \textbf{Data Fields (refer to Table~\ref{table1}):} This includes information on self-disclosed direct and indirect emissions, self-disclosed direct and indirect energy consumption, specific emission factors relevant to the industry, as well as geographic location and coordinates of company's subsidiaries.
    
    \item \textbf{Data Sources:} The information is compiled from annual reports, ESG reports, and other relevant environmental reports. These are sourced from major stock exchanges, including the Shanghai Stock Exchange (SSE) and Shenzhen Stock Exchange (SZSE).
    
    \item \textbf{Time Period: January 2019 to January 2023}.
    
    \item \textbf{Frequency: Annual data}.
\end{itemize}

\subsection{National/Regional Power Generation Emission Data}\label{subsection2}

This section details the methodology employed to collect and analyze national/regional power generation emission data, which helps estimate and validate a company's high-resolution and actual Scope 2 emissions. Among all diverse sources of company-level emissions, the emissions from the electricity consumed by a company play a critical role, encapsulated under the Scope 2 emissions category. Overall, we designed four steps for collecting and processing raw data.

\begin{enumerate}
\item \textbf{Data Retrieval}: We systematically compiled an exhaustive dataset encompassing the period spanning from January 1, 2019, to January 1, 2023, comprising critical variables such as installed capacity, fuel classifications, fuel consumption patterns, and cumulative generation hours for the entirety of China's power plants. This comprehensive assemblage of data was acquired from authoritative sources, including the official platforms or publicly accessible documents of the Global Power Plant Database, the annual reports of the China Electricity Council, the IPCC Emission Factor Database, and the authoritative emissions and power generation disclosures released by major power corporations in China. The collection of the dataset from diverse and reliable sources constitutes the fundamental bedrock upon which the estimation and verification of Scope 2 emissions are predicated.

\item \textbf{Origins and Power Mix Profiles for Each Company}: In the computation of a company's Scope 2 emissions attributed to electricity consumption, an initial step involves the identification of the geographic district from which the electricity is sourced. Subsequently, an analysis of the power generation composition linked to the company's electricity utilization is undertaken. The generation of electricity is predominantly executed within power plants, each characterized by a distinctive emissions profile shaped by variables such as fuel category (e.g., coal, natural gas, renewables), technological modalities, operational efficiency, and other pertinent parameters (refer to Table~\ref{table2}). 

\item \textbf{Emission Calculation for Power Plants}: The carbon emissions from power plants encompass three distinct elements: emissions arising from fossil fuel combustion within the power generation process, emissions associated with purchased electricity from external sources, and emissions resulting from the desulfurization process. In the context of desulfurization, the fundamental principle involves employing carbonate aqueous mixtures to facilitate a displacement reaction with sulfur dioxide generated post-coal combustion. This chemical interaction yields sulfites and carbon dioxide as byproducts. This fraction of carbon dioxide emissions can constitute around 20-30\% of the overall carbon emissions attributed to a coal-fired unit, signifying a substantial factor in comprehensive carbon accounting. The majority of thermal power units employ either gypsum or limestone desulfurization techniques. Notably, the IPCC accounting methodology for power facilities encompasses the quantification of carbon dioxide emissions stemming from the desulfurization process. 

Thus, the total carbon emissions E (in tons) for a power generation company can be defined as a mathematical formula that takes into account all the elements:

\[
E = E_{\text{fuel}} + E_{\text{desulfurization}} + E_{\text{electricity}}
\]
Each item's emissions can generally be summarized as ``activity level × carbon emission intensit'', which is:

\begin{align*}
E_{\text{fuel}} & = AD_{\text{fuel}} \times EF_{\text{fuel}} \\
E_{\text{desulfurization}} & = AD_{\text{desulfurization}} \times EF_{\text{desulfurization}} \\
E_{\text{electricity}} & = AD_{\text{electricity}} \times EF_{\text{electricity}}
\end{align*}

Here, \(AD_{\text{fuel}}\), \(AD_{\text{desulfurization}}\), and \(AD_{\text{electricity}}\) correspond to the ``activity levels'' of each production activity respectively, i.e., the heat released during fuel combustion, the weight or volume of carbonate consumed during desulfurization, and the plant's electricity consumption. \(EF_{\text{fuel}}\), \(EF_{\text{desulfurization}}\), and \(EF_{\text{electricity}}\) correspond to the ``carbon emission intensity'' of each production activity, i.e., the greenhouse gas carbon equivalent released per joule during fuel combustion, the carbon dioxide mass released per unit carbonate displacement reaction during desulfurization, and the carbon emissions per kWh for plant electricity. In actual calculations, the ``activity level'' corresponding to the production activity is often directly or indirectly calculated from the raw material consumption during the company's production operation, and the ``carbon emission intensity'' corresponding to the production activity is usually a constant obtained from a large amount of statistics in the corresponding field, either theoretical or experimental. 

\item \textbf{Emission Calculation for Regional Grids}: The grid emission intensity factor stands as a pivotal metric in the assessment of the carbon implications of electrical consumption, quantifying greenhouse gas emissions per unit of electricity generated. To calculate this metric, an amalgamation of data from all power plants within the grid is imperative. This data encompasses intricate details encompassing fuel categorization, consumption, and corresponding electricity output. The summative emissions emanating from each power plant can be ascertained by delving into the emissions attributes intrinsic to specific fuel categories, such as coal or renewables. By aggregating emissions across the entirety of power plants and subsequently dividing them by the cumulative electricity generated, the grid emission intensity factor is established. Recognizing the substantial variation in the power mix—reflecting the diversity of electricity sources across regions and its temporal evolution—we maintain a consistent regimen of data updates and recalibration. This diligence is pivotal to upholding the precision and reliability of our analytical results.

\end{enumerate}

With the aforementioned methods, we gathered six categories of national/regional power generation emission data. Each category is characterized by its field name, data source, and update frequency as detailed below:
\begin{itemize}
    \item \textbf{Data Fields (Refer to Table~\ref{table2}):}
    \begin{itemize}
        \item \textbf{Unit Fuel Type}: The type of fuel used in the power generation unit.
        \item \textbf{Fuel Carbon Emission Factor}: The amount of carbon emissions produced per unit of fuel consumed.
        \item \textbf{Installed Capacity of the Unit}: The maximum power output that the unit can produce.
        \item \textbf{Number of Hours of Electricity Generation Output}: The total hours of electricity generation and the corresponding power output in a year. 
        \item \textbf{Amount of Fuel Consumed per Unit of Electricity}: The fuel consumption rate for producing a unit of electricity.
        \item \textbf{Carbon Emissions of the Power Generation Company}: The total carbon emissions produced by the power generation company.
        \item \textbf{Carbon Emission Factor of the Power Generation company}: The emission factor specific to the power generation company.
        \item \textbf{Regional Scope 2 Carbon Emissions}: The indirect carbon emissions within the region.
        \item \textbf{Regional Scope 2 Carbon Emission Factor}: The emission factor specific to the indirect emissions within the region.
    \end{itemize}
    \item \textbf{Data Sources:}
    \begin{itemize}
        \item Global Power Plant Database
        \item China Electricity Council Annual Report 
        \item IPCC Emission Factor Database
        \item Power generation and emission data released by major power companies in China
        \item National standard GB21258-2017
        \item Company Greenhouse Gas Emission Accounting Method and Reporting Guide - Power Generation Facilities
        \item Shanghai Stock Exchange
        \item Shenzhen Stock Exchange
        \item Hong Kong Stock Exchange Annual Reports
        \item \textbf{Time Period: January 2019 to January 2023}
        \item \textbf{Time Frequency: Data is available on an annual, quarterly, and monthly basis.}

    \end{itemize}
 \end{itemize}

\begin{table}[!h]
    \centering
    \begin{tabularx}{\linewidth}{|l|X|}
        \hline
        \textbf{Data Type} & \textbf{Specific Data} \\
        \hline
        Unit Fuel Type and Carbon Emission Factor & Coal, unit: tons/TJ \\
        & Natural gas, unit: tons/TJ \\
        & Non-conventional fuels: coal gangue, coal slurry, coal water slurry, etc., unit: tons/TJ \\
        \hline
        Unit Installed Capacity & Maximum output power of the unit, unit: MW \\
        \hline
        Generation Hours/Power Generation & Annual generation hours for company, province, region, country, unit: h \\
        & Annual power generation for company, province, region, country, unit: MWh \\
        \hline
        Fuel Consumption per kWh & Standard coal consumption per kWh, unit: kg/kWh \\
        & Natural gas consumption per kWh, unit: m\textsuperscript{3}/kWh \\
        \hline
        Power Generation company Carbon Emission Data & Carbon emission, unit: tons \\
        & Carbon emission factor, unit: tons/MWh \\
        \hline
        Regional Scope 2 Carbon Emission Data & Carbon emission, unit: tons \\
        & Carbon emission factor, unit: tons/MWh \\
        \hline
    \end{tabularx}
    \caption{Data Specification for National/Regional Power Generation Emission Data}\label{table2}
\end{table}

\subsection{Regional Multi-source Satellite Remote Sensing Data}\label{subsection3}

To facilitate the monitoring of Scope 1 carbon emissions for companies, a comprehensive dataset was compiled, encompassing diverse sources such as multi-source satellite data, environmental data, CEMs data, and supplementary data associated with company infrastructures that produce emissions. These data sets are independent of the self-disclosed data by companies and, by leveraging the devised carbon emission estimation model (detailed in Section \ref{scope1estimation}), enable a higher temporal and spatial resolution for quantifying Scope 1 carbon emissions. As a result, we are capable of verifying the reported Scope 1 emissions from chosen companies, while also providing estimations for Scope 1 emissions from undisclosed entities. In summary, we propose a two-step approach for the collection and pre-processing of data.

\begin{enumerate}
\item \textbf{Data Retrieval}: We gathered multi-source satellite remote sensing data, including the column-averaged dry air mole fraction of carbon dioxide (XCO$_2$) and related information, as well as high-resolution satellite imagery. The XCO2 and related data were sourced from Level-2 products of OCO-2, OCO-3, and TanSat carbon satellites. The high-resolution satellite imagery was acquired from Level-2 data of Landsat 8/9 and Sentinel-2 satellites. For improving the estimation of Scope 1 carbon emissions using satellite data, ERA-5 environmental data were acquired from the European Centre for Medium-Range Weather Forecasts (ECMWF). Furthermore, we obtained power plant data equipped with Continuous Emissions Monitoring Systems (CEMs) from the U.S. Environmental Protection Agency (EPA). This dataset encompasses both carbon emission data and electricity consumption data monitored by CEMS, rendering it suitable for constructing the dataset used in carbon emission estimation models building. Finally, we collected data on emission-producing infrastructures owned by various companies from Global Energy Monitor (GEM). In the data retrieval, we collected raw data from various modalities.

\item \textbf{Key Information Extraction}: To construct a carbon emission AI model for estimating Scope 1 carbon emissions of companies, it is imperative to initially extract relevant portions of data from the multi-modal data set. Concerning the carbon satellite data, we extracted XCO$_2$, Area coordinates, Time, XCO$_2$ uncertainty, and XCO$_2$ quality label information. For Landsat 8/9 and Sentinel-2 satellites, we extracted the Red, Green, and Blue bands satellite data to obtain RGB satellite imagery. From Continuous Emissions Monitoring Systems (CEMS) power plant data, we retrieved hourly carbon emission data and electricity generation data for the power plants. Lastly, leveraging NLP algorithms and manual validation, we matched China's publicly listed companies with their corresponding emission-producing infrastructures within the Global Energy Monitor (GEM) data set  (refer to Table~\ref{table3}). 
\end{enumerate}

\begin{itemize}
    \item \textbf{Data Fields (Refer to Table~\ref{table3}):}
    This includes carbon satellite records, satellite imagery, environmental variables, CEMS records, and emission-producing infrastructure information.
    \begin{itemize}
        \item \textbf{Carbon Satellites Data}: The important records of carbon satellites, including column-averaged dry air mole fraction of carbon dioxide (XCO$_2$), area coordinates, time, XCO$_2$ uncertainty, and XCO$_2$ quality label.
        \item \textbf{Satellite Imagery}: RGB images of Earth captured by orbiting satellites, which provide detailed visual information about geographical and environmental features.
        \item \textbf{Environmental Variables}: Temperature, u/v wind speeds, and pressure.
        \item \textbf{CEMs Data}: The data recorded from power plants equipped by CEMs.
        \item \textbf{Emission-producing Infrastructures Information}: The information on emission-producing infrastructures, especially the owner companies and the coordinates.
    \end{itemize}

\begin{table}[!h]
\centering
\begin{tabularx}{\textwidth}{lX}
\toprule
\textbf{Data Type} & \textbf{Specific Data} \\
\midrule
\multirow{5}{*}{Carbon satellites data} & Column-averaged dry air mole fraction of carbon dioxide (XCO$_2$): The column-averaged concentration of carbon dioxide in a specific area (in units of ppm) \\
 & Area coordinates: The geographical coordinates where the measurement was taken, including the longitude and latitude of the center of the sounding field-of-view. Unit: latitude and longitude degrees. \\
 & Time: The exact time of the sounding in seconds. Unit: s.\\
 & XCO$_2$ Uncertainty: The level of uncertainty or margin of error associated with the XCO$_2$ data. \\
 & XCO$_2$ Quality Label: A label indicating the quality or reliability of the XCO$_2$ data. \\
\midrule
\multirow{1}{*}{Satellite imagery} & RGB images of Earth captured by orbiting satellites. \\
\midrule
\multirow{4}{*}{Environmental variables} & Temperature:  Temperature of air at 2m above the earth surface. Unit: K.\\
 & u/v wind speeds: Eastward (u) and Northward (v) component of the 10m wind. u is the horizontal speed of air moving towards the east. v is the horizontal speed of air moving towards the north. Unit: m/s. \\
 & Pressure:  Pressure (force per unit area) of the atmosphere on the earth surface. Unit: Pa. \\
\midrule
\multirow{2}{*}{CEMs data} & CEMs emissions: CEMs measured Scope 1 emissions of power plants. Unit: tons. \\
 & Gross load: hourly power generation of power plants. Unit: MWh. \\
\midrule
\multirow{2}{*}{Emission-producing infrastructures information} & Infrastructures' coordinates: The longitude and latitude of  infrastructures. Unit: latitude and longitude degrees. \\
 & Owner: The owner companies of infrastructures. \\
\bottomrule
\end{tabularx}
\caption{Data Type of Regional Multi-source Satellite Remote Sensing Data}\label{table3}
\end{table}

    \item \textbf{Data Sources:}
    \begin{itemize}
        \item \textbf{Orbiting Carbon Observatory-2 (OCO-2)}: A satellite designed to monitor global carbon dioxide levels.
        \item \textbf{Orbiting Carbon Observatory-3 (OCO-3)}: A subsequent version of the OCO-2, with enhanced capabilities.
        \item \textbf{Global Carbon Dioxide Monitoring Science Experimental Satellite (Tansat)}: A satellite specifically focused on experimental monitoring of global carbon dioxide levels.
        \item \textbf{USGS Landsat 8/9 Level 2 (Landsat 8/9)}: a satellite program by the U.S. Geological Survey that provides high-resolution, multispectral imagery of the Earth's surface, facilitating environmental monitoring and land change detection
        \item \textbf{Sentinel-2 MultiSpectral Instrument, Level-2A (Sentinel-2)}: A high-resolution optical instrument aboard the Sentinel-2 satellite, designed for detailed Earth observation in multiple spectral bands.
        \item \textbf{ERA5 Hourly Data on Single Levels from 1940 to Present (ERA-5)}: A high-resolution, time-consistent atmospheric reanalysis dataset, providing hourly updates on various atmospheric variables spanning from 1940 to the current date.
        \item \textbf{CEMs Data}: The CEMs emissions data are available from the EPA Clean Air Markets Program Data.
        \item \textbf{ERA5 Hourly Data on Single Levels from 1940 to Present (ERA-5)}: A high-resolution, time-consistent atmospheric reanalysis dataset, providing hourly updates on various atmospheric variables spanning from 1940 to the current date.
        \item \textbf{Time Period: October 2014 to January 2023} 
        \item \textbf{Time Frequency: Satellites orbit the Earth while conducting scans, acquiring scanning area data in near-real time.} 
    \end{itemize}
\end{itemize}


\section{Methodology}\label{sec3}
\subsection{Scope 1 Emission Estimation: Carbon Emission AI Estimation Model}\label{scope1estimation}
\begin{figure*}[!t]
   \begin{center}
   \includegraphics[width=\linewidth]{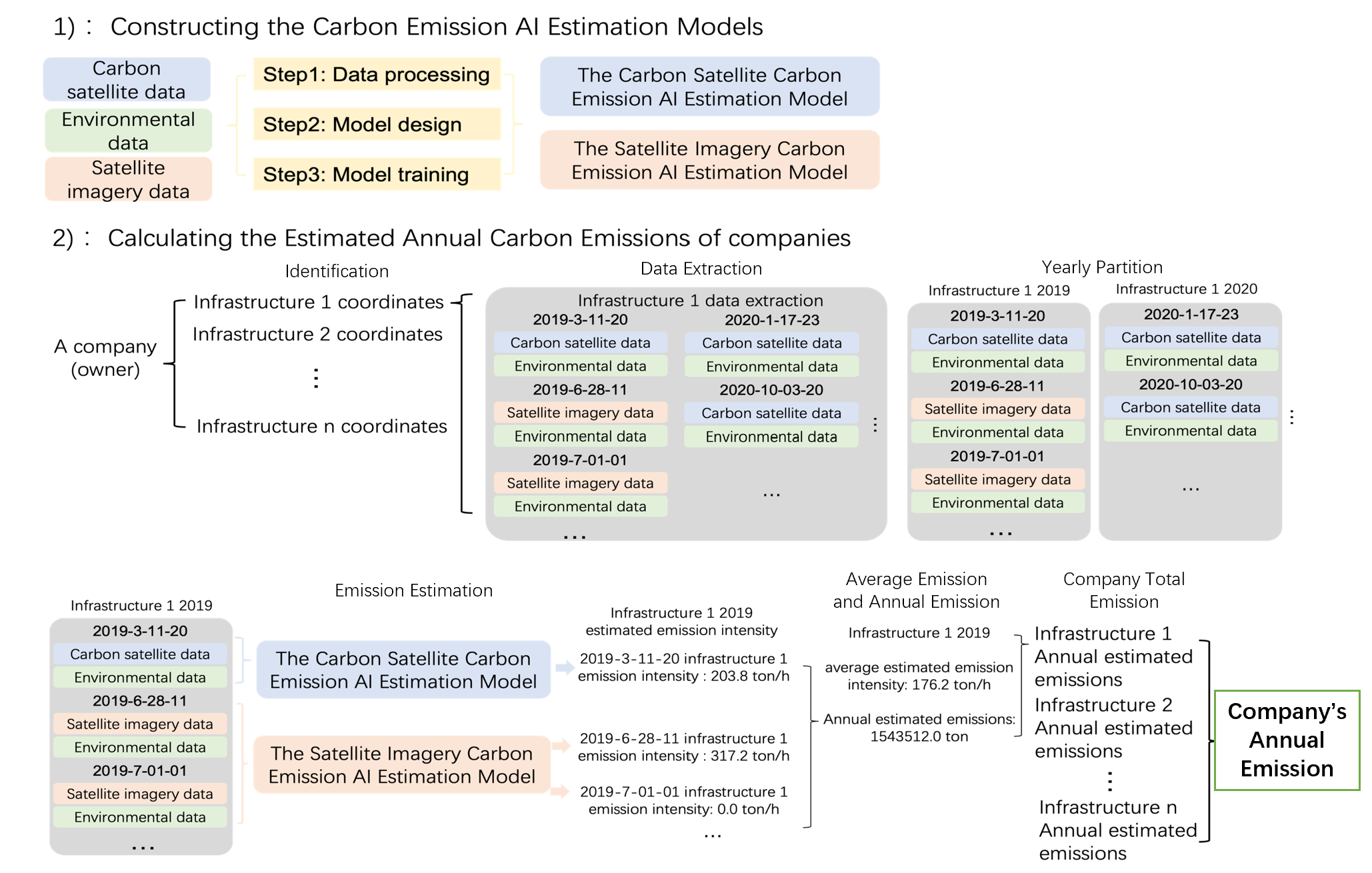}
   \end{center}
      \caption{The Two-step Framework of Scope 1 Emission Estimation: The first step is constructing the Carbon Emission AI Estimation Models. The second step is calculating the estimated annual carbon emissions of companies.}
   \label{scope1}
   \end{figure*}

As shown in Figure \ref{scope1}, estimating Scope 1 carbon emissions for companies involves two steps: (1) Constructing the Carbon Emission AI Estimation Models. (2) Calculating the estimated annual carbon emissions of companies. After estimating the annual carbon emissions of all company-owned infrastructures, summing the annual carbon emissions of all infrastructures owned by the company to obtain the annual carbon emissions for the entire company. 

Our carbon emission AI estimation models for direct emission estimation can be classified into two categories: (1) The Carbon Satellite Carbon Emission AI Estimation Model, which utilizes carbon dioxide concentration data (XCO2) from carbon satellites and environmental data to estimate the regional carbon emission intensity; (2) The Satellite Imagery Carbon Emission AI Estimation Model, on the other hand, relies on high-resolution satellite imagery data (depicting carbon-emitting plumes) and environmental data to estimate carbon emission intensity from facilities such as power plants and factories. The construction methodologies for both models are analogous, comprising three phases: data processing, model design, and model training.

\subsubsection{The Carbon Satellite Carbon Emission AI Estimation Model}
The Carbon Satellite Carbon Emission AI Estimation Model utilizes carbon satellite, electricity, and environmental data. We designed three steps for predicting high-resolution regional carbon emissions in which companies are located: 
\begin{itemize}
\item \textbf{Step 1: Data Processing}
\begin{enumerate}
    \item \textbf{Matching based on Location and Time:} We merged carbon sources within a \(5km^2\) area, matched the surrounding area's carbon satellite data according to the carbon source location, divided satellite stripe data at different times, and deleted carbon star data that is more than 40km away from the carbon source.
    \item \textbf{Filtering based on Statistical Information:} We filtered out outliers based on the mean, standard deviation, minimum, maximum, number of power plants in the area, and number of records in the satellite stripe data. A large number of data without carbon emission labels remained, suitable for large-scale self-supervised pre-training.
    \item \textbf{Filtering based on the Plume Model:} We then selected carbon satellite stripe data with measured or calculated carbon emission data and chose satellite stripe data facing the wind direction relative to the carbon source based on wind speed. The carbon satellite stripe data has a plume in the direction of the wind, which should not be overly complex. After filtering based on the plume model, a small amount of data with carbon emission labels was obtained for small-scale supervised learning.
\end{enumerate}

\item \textbf{Step 2: Model Design}
\begin{enumerate}
    \item \textbf{Data Encoding:} We encoded carbon satellite, electricity, and environmental data into a format suitable for the model. This includes encoding regional latitude and longitude, regional carbon dioxide mixture ratio, wind speed in the regional longitude and latitude direction, and statistical values of the satellite stripe into vector form.
    \item \textbf{Model Structure:} We used the CarbonNet structure proposed in \cite{zhang2022near} as shown in Figure \ref{carbonnet}. The model architecture employs the classical BERT network structure, with its core being the Transformer architecture characterized by attention mechanisms and parallel processing capabilities. The Transformer architecture usually comprises an encoder and a decoder; however, in the case of BERT, only the encoder component was utilized for pre-training. This structure can effectively capture the associative information present in the XCO$_2$ data from different regions of carbon satellites, generating high-dimensional feature representations.
\end{enumerate}

\begin{figure*}[!t]
   \begin{center}
   \includegraphics[width=\linewidth]{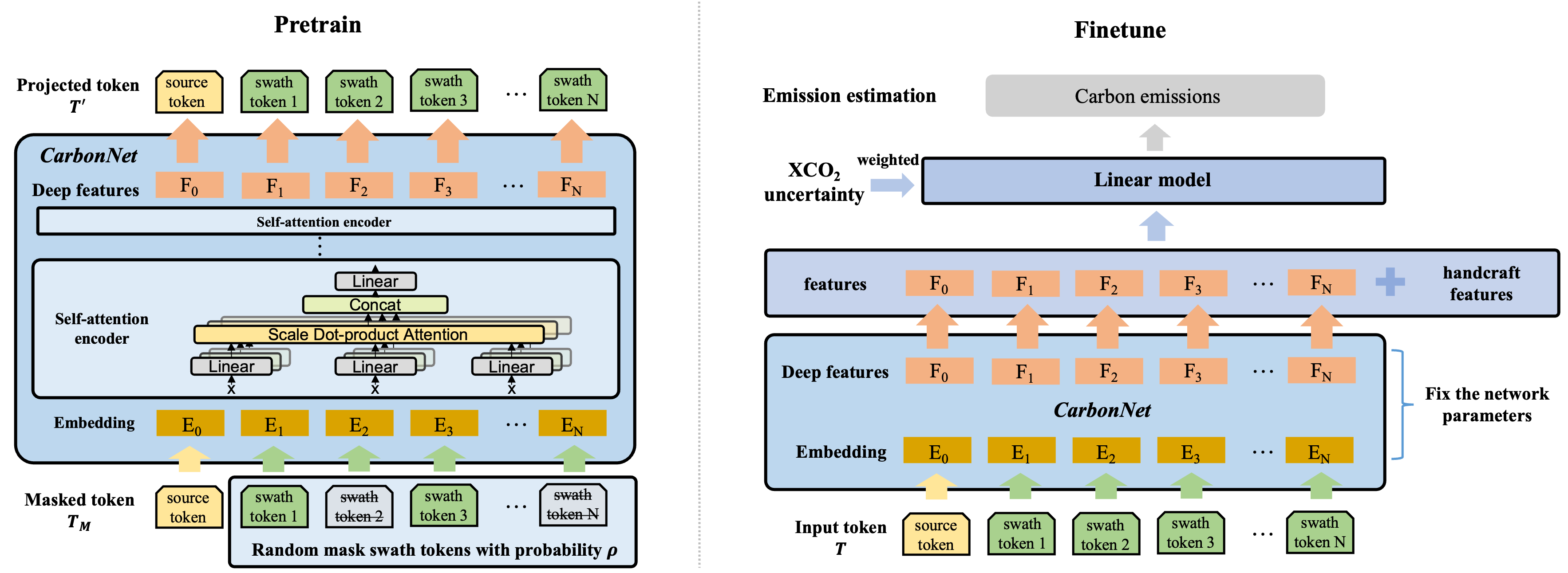}
   \end{center}
      \caption{Self-supervised Pre-training Stage and Supervised Learning Finetune Stage \cite{zhang2022near}.}
   \label{carbonnet}
   \end{figure*}

\item\textbf{Step3: Model Training}
\begin{enumerate}
    \item \textbf{Self-supervised Pre-training based on Large-scale Non-carbon Emission Labeled Data:} A deep neural network model based on the self-attention mechanism was constructed, as well as a self-supervised pre-training method based on large-scale non-carbon emission label data. During the training of the network model using the self-supervised pre-training method, 25\% of the input carbon satellite stripe data was randomly masked, and the model's output label was the complete carbon satellite stripe data. This training method requires the network to recover the complete carbon satellite stripe data when a portion of the carbon satellite stripe data is missing. The trained model exhibits good feature representation capability, extracting network features as part of the input feature for subsequent supervised learning.
    \item \textbf{Supervised Learning based on Small-scale Carbon Emission Labeled Data:} we then input carbon satellite data into the pre-trained deep neural network model based on the self-attention mechanism to extract features, and combined these with manually extracted features (including carbon source location, wind speed at the carbon source location, and the mean and standard deviation of carbon satellite stripe data). A regression model was used for supervised learning on small-scale data with carbon emission labels, establishing a mapping model from multi-modal data (carbon satellite data and other data) to carbon source carbon emission data.
\end{enumerate}
\end{itemize}

\subsubsection{The Satellite Imagery Carbon Emission AI Estimation Model}
\begin{itemize}
\item \textbf{Step 1: Data Processing}
\begin{enumerate}
    \item \textbf{Matching based on Location and Time:} High-resolution satellite imagery data surrounding carbon source locations were matched for all time instances, and the matched data was subjected to location correction.
    \item \textbf{Carbon Source Plume Detection:} Carbon source plumes were detected within the high-resolution satellite imagery data, identifying records capturing carbon emission information.
    \item \textbf{Dataset Construction:} The construction of a dataset involves compiling high-resolution satellite imagery data corresponding to carbon source locations - creating a database of satellite imagery data associated with carbon emissions.
\end{enumerate}

\item\textbf{Step 2: Model Design}
\begin{enumerate}
    \item \textbf{Data Encoding:} Satellite imagery data and environmental data were stacked according to their respective locations and encoded as three-dimensional input for the model.
    \item \textbf{Model Structure:} The model architecture chosen employed the classical VIT network structure. The core idea of the VIT model is to partition the image into a sequence of image patches, which are then transformed into sequential data, facilitating the application of the Transformer's self-attention mechanism. Each image patch, after undergoing a linear transformation, was considered as a part of the input sequence, subsequently entering the Transformer encoder. This approach enables VIT to capture relationships between different regions, comprehensively understanding the global context.
\end{enumerate}

\item\textbf{Step 3: Model Training}
\begin{enumerate}
    \item \textbf{Pre-train:} The model was first pre-trained on a large-scale dataset of remote sensing imagery using self-supervised learning methods. Self-supervised tasks, such as image rotation and occlusion restoration, enabled the model to learn information from remote-sensing images.
    \item \textbf{Finetune:} The pre-trained VIT model was fine-tuned using the processed data to estimate carbon emissions, ensuring it adapted for the specific task.
\end{enumerate}
\end{itemize}

\subsubsection{Calculating the Estimated Annual Carbon Emissions of Companies}
For a given company, our methodology to calculate the annual carbon emission of its industrial factories and other subsidiaries is composed of eight steps, as shown in Figure \ref{scope1}:
\begin{itemize}
    \item \textbf{Identification:} Determine names and coordinates of all emission-producing infrastructures linked to the company.
    \item \textbf{Data Extraction:} Extract multi-source satellite and relevant environmental data based on the coordinates of identified infrastructures.
    \item \textbf{Yearly Partition:} Segment the processed data according to specific years.
    \item \textbf{Emission Estimation:} Use both The Carbon Satellite Carbon Emission AI Estimation Model and The Satellite Imagery Carbon Emission AI Estimation Model to calculate the Scope 1 carbon emission intensity (unit: ton/h) during specified time frames.
    \item \textbf{Annual Average:} Determine the yearly average Scope 1 carbon emission intensity by averaging the estimated values across all infrastructures.
    \item \textbf{Annual Emissions:} Multiply the annual average emission intensity by the total number of hours in a year to deduce annual Scope 1 emissions for infrastructures.
    \item \textbf{Company Total Emission:} Sum the annual Scope 1 emission from all infrastructures to calculate the total annual Scope 1 emission for the company.
    \item \textbf{Scaling:} Apply the procedure across all publicly listed companies to determine their respective annual Scope 1 carbon emissions.
\end{itemize}


\subsection{Scope 2 Emissions Estimation:  Carbon Accounting based on the High-resolution Power Emission Factors}
Within the ambit of the top-down carbon dioxide emissions estimation paradigm, the determination of Scope 2 emissions transpires through the multiplication of consumed electricity by the national grid emission factor\cite{eggleston20062006}. However, a notable verity surfaces: the grid emission factors are distinctly heterogeneous across divergent companies and regions. To facilitate the estimation of a company's Scope 2 emissions, two distinct approaches emerge: the first entails the utilization of precise regional or provincial grid emission factors to compute emissions, while the second mandates the application of the China Electricity Council Annual Report's generation and transmission data, the fuel emission factor data enshrined in the IPCC Emission Factor Database, and the fuel consumption per kilowatt-hour data gleaned from the standardized document DB 33/644—2012(2013) to compute the average carbon emission factor within the specified locale. Despite the methodological intricacies inherent in the second approach, it resolves the intricacies associated with regional or provincial emission factor calculations.
\begin{itemize}
\item\textbf{Database and Methodology}: The estimation of Scope 2 emissions relies on multiple industrial reports and standard documents, including but not limited to:
\begin{enumerate}
\item \textbf{China Electricity Council Annual Report}: It provides the generation and transmission data provincially by year.
\item \textbf{IPCC Emission Factor Database}: It provides the emission factor data of different types of coal and gas by year.
\item \textbf{DB 33/644—2012(2013)}: This is a standard document recording the fuel consumption per kilowatt-hour data for different types of thermal generators.
\end{enumerate}
The emission factor $EF_{\text{region}}$ calculation method is shown by the equation below

\begin{align*}
EF_{\text{region}} & = \frac{(E_{\text{gen}}-E_{\text{out}})\times c \times EF_{\text{fuel}}+E_{\text{in}}\times EF_{\text{in}}}{E_{\text{gen}}-E_{\text{out}}+E_{\text{in}}}
\end{align*}

where $E_{\text{gen}}$, $E_{\text{in}}$, and $E_{\text{out}}$ are the electricity energy generated, transmitted-in, and transmitted-out of the certain region, $c$ is the fuel consumption per kilowatt-hour, and $EF_{\text{fuel}}$ and $EF_{\text{in}}$ are the emission factors of the fuel (tone/TJ) and input energy (tone/MWh) respectively.
\end{itemize}

\begin{itemize}
\item\textbf{Benefits and Applications}: Through the amalgamation of extensive and diverse data reservoirs, this approach proffers a resilient and dependable avenue for the estimation and validation of Scope 2 carbon emissions. In doing so, it delineates a substantial stride within the realm of environmental surveillance and the company expanse of carbon quantification.
\end{itemize}


\subsection{Scope 3 Emission Estimation using Company Equity Penetration Data }
Estimating the Scope 3 carbon emissions using company equity penetration data involves considering the indirect emissions associated with a company's value chain and its ownership proportion in various assets or operations. Scope 3 emissions cover all indirect emissions in a company's value chain, excluding those in Scope 1 (direct emissions) and Scope 2 (energy indirect emissions). This includes emissions from upstream production, transportation, end-use of sold products, and investments.
We use four steps to employ equity penetration data for Scope 3 emission estimation:
\begin{enumerate}
    \item \textbf{Identify Ownership Stakes}: Extract a company's percentage of equity penetration in different assets, operations, or joint ventures.
    \item \textbf{Scope 3 Emission Sources}: For each asset, operation, or joint venture identified, ascertain the potential sources of Scope 3 emissions, ranging from raw material sourcing to product distribution and end-of-life treatment.
    \item \textbf{Proportional Emission Calculation}: For each identified Scope 3 emission source, multiply the company's ownership percentage by the total carbon emissions associated with that source. E.g., if a company owns 30\% equity in an operation responsible for 50,000 tCO2 in transportation-related emissions, the company's proportional emissions would be 15,000 tCO2.
    \item \textbf{Aggregate All Emissions}: Sum up the proportional Scope 3 emissions from all relevant assets and operations to get a total figure for the company's equity-adjusted Scope 3 emissions.    
    \item \textbf{Validation \& Cross-reference}: Ensure the derived Scope 3 emission figures are validated. Cross-referencing with other data sources, third-party audits, or using estimation methodologies can enhance the accuracy and credibility of the figures.     
\end{enumerate}
By adeptly using equity penetration data with other relevant data sources and methodologies, our database can provide a robust and holistic estimate of a company's Scope 3 emissions, shedding light on its complete carbon footprint across the value chain.

\section{Data Records}
High-Resolution Carbon Emissions Database for China’s Listed Companies, derived from carbon satellite and power data, is available through the web page (\url{intelligentcarbon.ai}). This comprehensive database provides users with detailed emissions data and other environmental information of China's A-share listed companies in Microsoft Excel formats. To ensure clarity and credibility, we have provided a thorough listing of primary source files. 
The variables within the data file are as follows.
\begin{itemize}
\item \textbf{company\_name}: the name of a listed company
\item \textbf{stock\_id}: the stock code of a listed company
\item \textbf{listing\_location}: the stock exchange where a company is listed (e.g., SHEX/SZEX)
\item \textbf{industry}: the industry that a listed company belongs to
\item \textbf{geographic\_coordinates}: geographic coordinates of a listed company
\item \textbf{geographical\_location} geographic location of a listed company
\item \textbf{revenue\_2022}: revenue of main business of a listed company in 2022 (in ten thousand CNY) 
\item \textbf{revenue\_2021}: revenue of main business of a listed company in 2021 (in ten thousand CNY) 
\item \textbf{revenue\_2020}: revenue of main business of a listed company in 2020 (in ten thousand CNY) 
\item \textbf{revenue\_2019}: revenue of main business of a listed company in 2019 (in ten thousand CNY) 
\item \textbf{total\_emission\_2022}: total carbon emissions of a listed company in 2022 (tCO2)
\item \textbf{total\_emission\_2021}: total carbon emissions of a listed company in 2021 (tCO2)
\item \textbf{total\_emission\_2020}: total carbon emissions of a listed company in 2020 (tCO2)
\item \textbf{total\_emission\_2019}: total carbon emissions of a listed company in 2019 (tCO2)
\item \textbf{emission\_intensity\_2022}: proportion of total carbon emissions to main business revenue of a listed company in 2022
\item \textbf{emission\_intensity\_2021}: proportion of total carbon emissions to main business revenue of a listed company in 2021
\item \textbf{emission\_intensity\_2020}: proportion of total carbon emissions to main business revenue of a listed company in 2020
\item \textbf{emission\_intensity\_2019}: proportion of total carbon emissions to main business revenue of a listed company in 2019
\item \textbf{emission\_mitigation}: reduction in emission intensity of a listed company from 2019 to 2022
\item \textbf{scope\_1\_emission}: annual Scope 1 carbon emissions of a listed company from 2019 to 2022
\item \textbf{scope\_2\_emission}: annual Scope 2 carbon emissions of a listed company from 2019 to 2022
\item \textbf{scope\_3\_emission}: annual Scope 3 carbon emissions of a listed company from 2019 to 2022
\item \textbf{renewable\_energy}: renewable energy purchase amount/proportion of renewable energy of a listed company
\item \textbf{emission\_coverage}: whether the self-disclosed Emission of a listed company covers the Entire Company subsidiaries.
\item \textbf{purchased\_ electricity}: purchased electricity of a listed company
\item \textbf{purchased\_ heat}: purchased thermal power of a listed company
\item \textbf{purchased\_ natural\_gas}: natural gas usage of a listed company \item \textbf{purchased\_ coal\_usage}: coal usage of a listed company
\item \textbf{purchased\_ gasoline\_usage}: gasoline usage of a listed company
\item \textbf{purchased\_ disel\_usage}: diesel usage of a listed company
\item \textbf{accounting\_description}: carbon accounting scope, methodology explanation, carbon emission factor explanation of a listed company
\item \textbf{certification\_description}: whether a listed company has data audit certification report for the self-disclosed data
\item \textbf{green\_finance}: carbon reduction loans and other green financial emission reductions of a listed company
\item \textbf{data\_center\_emission}: company data center emissions of a listed company
\item \textbf{dual\_carbon\_target}: ``Carbon Neutrality, Carbon Peak" objectives of a listed company
\item \textbf{emission\_reduction\_timeline}: timeline of emission reduction plan for a listed company
\item \textbf{emission\_reduction\_target}: planned emission reduction amount of a listed company
\item \textbf{emission\_reduction\_tech}: emission reduction technologies of a listed company
\end{itemize}

\section {Data Validation}
\subsection{Cross-validation for Scope 1 Emissions}\label{subsec3}
The \textit{IPCC 2006 Guidelines for National Greenhouse Gas Inventories}, revised in 2019, advocate a top-down approach to estimate carbon dioxide emissions. However, the top-down approach for estimating company-level emissions faces several challenges: 1) top-down strategies derive from broader regional or national data to inform localized predictions, potentially missing company-specific variations; 2) Top-down models may miss out on capturing certain emission types and cannot represent short-term fluctuations in emissions. Hence, based on the traditional top-down approach, our database offers a comprehensive and more accurate view of emissions by integrating diverse data sources and continuous monitoring systems and leveraging satellite remote sensing measurements to validate the bottom-up national greenhouse gas inventory method. 

The database utilizes a constellation of multiple remote sensing satellites, with a particular emphasis on carbon-sensing satellites and multi-modal datasets. This resource is employed to develop tailored carbon emission estimation models for various industries, including but not limited to the electricity, steel, and cement sectors. These models facilitate the estimation of carbon emissions for companies within these industries. Subsequently, the derived dataset is employed to validate the reported carbon emissions within the Scope of disclosed company activities. Furthermore, the database fills the gap by providing supplementary carbon emission data for corporations without divulging their emission information. Given the broad scope, repeatable observations, objectivity, and uniformity of remote sensing data, it is suitable for cross-validating the direct emissions disclosed by companies.

\begin{itemize}
\item\textbf{Database and Methodology:} The cross-validation of Scope 1 emissions relies on multiple remote sensing satellites, including but not limited to:
\begin{enumerate}
\item \textbf{Orbiting Carbon Observatory-2 (OCO-2)}: It is specialized in monitoring global carbon dioxide levels.
\item \textbf{Orbiting Carbon Observatory-3 (OCO-3)}:An advanced version of OCO-2 with enhanced capabilities.
\item \textbf{TanSat (Global Carbon Dioxide Monitoring Satellite)}: Focused on experimental monitoring of global carbon dioxide levels.
\item \textbf{USGS Landsat 8/9 Level 2 (Landsat 8/9)}: A satellite program by the U.S. Geological Survey that provides high-resolution, multispectral imagery of the Earth's surface, facilitating environmental monitoring and land change detection
\item \textbf{Sentinel-2 MultiSpectral Instrument, Level-2A (Sentinel-2)}: A high-resolution optical instrument aboard the Sentinel-2 satellite, designed for detailed Earth observation in multiple spectral bands.
\end{enumerate}
\end{itemize}

Two separate models are deployed for validation: The Carbon Satellite Carbon Emission AI Estimation Model and The Satellite Imagery Carbon Emission AI Estimation Model. These models enable us to estimate the Scope 1 carbon emission intensity for infrastructures owned by companies and consequently calculate their Scope 1 carbon emissions. Overall, there are three steps for the cross-validation of Scope 1 emissions:
\begin{itemize}
\item \textbf{Scope 1 Emission Estimation}: The Carbon Satellite Carbon Emission AI Estimation Model and the Satellite Imagery AI Estimation Model were employed to estimate the carbon emission intensity (tons per hour) of the collected power plants or factories. The estimated carbon emission intensity was then summed and averaged annually to obtain the yearly average emission intensity for each power plant or factory. Subsequently, the annual carbon emissions (in tons) were calculated.

\item \textbf{Bias Adjustment}:  We established threshold rules to perform cross-validation between annually estimated Scope 1 carbon emissions derived from independent multi-source satellite data and the publicly disclosed Scope 1 carbon emissions of listed companies. If the deviation between the publicly disclosed Scope 1 carbon emissions and the annually estimated Scope 1 carbon emissions exceeds the threshold $T$, it is identified that there is an issue with the publicly disclosed Scope 1 carbon emissions of the listed company. In such cases, the database substitutes the publicly disclosed Scope 1 carbon emissions with the annually estimated Scope 1 carbon emissions and marks the company for that specific year. If the deviation is below the threshold $T$, the disclosed Scope 1 carbon emissions of the company are deemed reliable.

\item\textbf{Missing Value Imputation}: The derived emission data from satellite measurements is not just for validation but also for filling data gaps. The satellite-based estimates offer a reliable supplementary data source for companies that have not divulged their emission information. Scope 1 carbon emissions that were not disclosed within the publicly listed companies' reports were supplemented based on the estimation. This has led to an increase in the number of publicly listed companies and sectors covered by the released carbon database.
\end{itemize}

\begin{itemize}
\item\textbf{Benefits and Applications:}
\begin{enumerate}
\item \textbf{Broad Scope}: The extensive reach of remote sensing data ensures comprehensive coverage.
\item \textbf{Repeatable Observations}: The data can be consistently replicated, enhancing reliability.
\item \textbf{Objectivity and Uniformity}: The unbiased and standardized nature of remote sensing data ensures consistency.
\item \textbf{Cross-Validation Capability}: The remote sensing data is suitable for cross-validating the direct emissions disclosed by companies, providing an additional layer of verification.
By integrating cutting-edge technology and a diverse array of satellite resources, this approach offers a robust and innovative method for carbon emissions estimation and validation. It represents a significant advancement in the field of environmental monitoring and company-level carbon accounting \cite{zhang2022near}.
\end{enumerate}
\end{itemize}

\subsection{Validation for Scope 2 Emissions}\label{subsec3}
The Scope 2 emissions of a company entity encapsulate a facet of indirect carbon emissions encompassing both electricity and heat consumption. However, prevailing methodologies predominantly address the electricity component, rendering the estimations inherently prone to inaccuracies. To enhance the precision of Scope 2 emission estimation, a comprehensive approach is adopted. By integrating the Combined Heat and Power (CHP) data sourced from the China Electricity Council Annual Report with the company's electricity consumption data, the thermal energy consumption of the company is approximated. This estimation is further refined through amalgamation with the heat emission factor delineated by the IPCC. A discerning criterion is introduced—manifested as a 20\% threshold—wherein, if the calculated carbon emission from thermal energy surpasses 20\% of the corresponding electricity-based emissions, the Scope 2 emission estimate is accordingly validated. This validation yields a recalibrated Scope 2 emission value, constituting the cumulative summation of both electricity and heat-related carbon emissions.

\section{Conclusion}

The High-Resolution Carbon Emissions Database for China’s Listed Companies stands as a beacon of innovation and progress in the realm of environmental monitoring and corporate sustainability. Addressing the urgent and expansive challenge of carbon emissions tracking, the database we've developed not only bridges significant gaps in data but also fosters transparency and accountability in China, one of the world's most industrialized nations. Several salient points underscore the significance and implications of our efforts:

\begin{itemize}

\item \textbf{Comprehensive Coverage}: The database, which spans across China's A-share listed companies, encapsulates a broad spectrum of data, from company-specific information such as revenue and location to detailed emissions statistics for multiple years. This granularity and breadth are poised to become a reliable and valuable information resource for stakeholders across the board.

\item \textbf{Cutting-Edge Methodology}: Our approach to emissions validation, especially for Scope 1 and Scope 2 emissions, represents a vanguard methodology in the field. By leveraging advanced artificial intelligence algorithms, satellite remote sensing measurements, and multi-modal datasets, we ensure a level of precision and verifiability of company-level emission data. This commitment to technological excellence augments the reliability and credibility of the data we present.

\item \textbf{Enhanced Transparency}: The provision of primary source files, comprehensive documentation, and thorough cross-validation mechanisms instills a new degree of transparency in carbon reporting. Companies, investors, and policymakers can now peruse and evaluate the carbon footprints of listed companies. This transparency catalyzes informed decision-making and stimulates the achievement of carbon neutrality goals.

\item \textbf{Bridging Information Gaps}: The database doesn't merely validate self-disclosed emissions; it also provides estimations for non-disclosed emissions. As a result, we offer a more complete and holistic view of the carbon footprint landscape of China's listed companies.

\item \textbf{Strategic Implications}: With the more accurate and high-resolution carbon accounting, companies can now benchmark their performance, strategize on emission reductions, and work towards aligning with China's environmental goals. Concurrently, investors can make environmentally informed choices, and regulators can have better oversight of the company's environmental landscape.

\end{itemize}

As the world grapples with the pressing imperatives of climate change, the High-Resolution Carbon Emissions Database for China’s Listed Companies will undoubtedly play a central role in navigating our collective journey toward a sustainable future. As we continue to refine, expand, and enhance our efforts, we remain steadfast in our commitment to fostering a greener, more transparent, and sustainable corporate landscape in China.

\section{Appendix}
\begin{figure}[!h]
    \centering
    \includegraphics[width=1\linewidth]{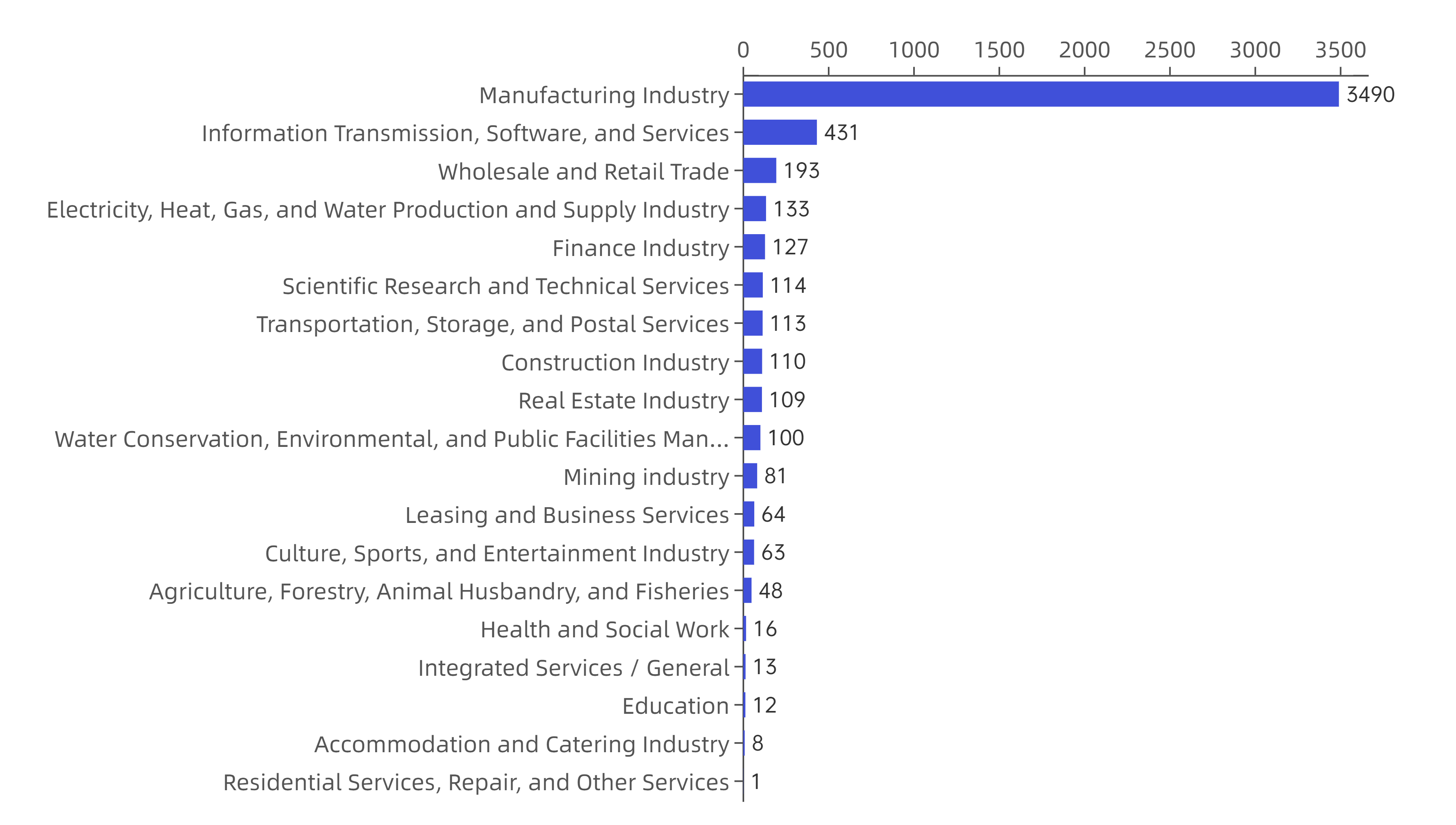}
    \caption{The Number of Companies of Industries Covered in the High-Resolution Carbon Emissions Database for China’s Listed Companies}
    \label{fig: database flow chart}
\end{figure}


\begin{thebibliography}{1}

\bibitem{christensen2015incentives}
Christensen, Hans B., Edward Lee, Martin Walker, and Cheng Zeng. "Incentives or standards: What determines accounting quality changes around IFRS adoption?." European Accounting Review 24, no. 1 (2015): 31-61.

\bibitem{kolk2008corporate}
Kolk, Ans, David Levy, and Jonatan Pinkse. "Corporate responses in an emerging climate regime: The institutionalization and commensuration of carbon disclosure." European accounting review 17, no. 4 (2008): 719-745.

\bibitem{cho2015organized}
Cho, Charles H., Matias Laine, Robin W. Roberts, and Michelle Rodrigue. "Organized hypocrisy, organizational façades, and sustainability reporting." Accounting, organizations and society 40 (2015): 78-94.

\bibitem{plumlee2015voluntary}
Plumlee, Marlene, Darrell Brown, Rachel M. Hayes, and R. Scott Marshall. "Voluntary environmental disclosure quality and firm value: Further evidence." Journal of accounting and public policy 34, no. 4 (2015): 336-361.

\bibitem{eggleston20062006}
Eggleston, H. S., Leandro Buendia, Kyoko Miwa, Todd Ngara, and Kiyoto Tanabe. "2006 IPCC guidelines for national greenhouse gas inventories." (2006).

\bibitem{zhang2022near}
Zhang, Zhengwen, Jinjin Gu, Junhua Zhao, Jianwei Huang, and Haifeng Wu. "Near Real-time CO$_2 $ Emissions Based on Carbon Satellite And Artificial Intelligence." arXiv preprint arXiv:2210.09850 (2022).

\end{thebibliography}
\end{document}